\documentclass[a4paper,12pt]{article}
\usepackage{amssymb}
\usepackage{graphicx}

\def\journal#1#2#3#4{{\it #1} {\bf #2} #3 #4}

\def\grad{\mathbf{\nabla}}

\def\v{\mathbf{v}}
\def\a{\mathbf{A}}
\def\f{\mathbf{F}}

\def\om{\mathbf{\omega}}
\def\l{{\cal L}}
\def\h{{\cal H}}
\def\f{{\cal F}}
\def\z{{\cal Z}}

\def\pd{\partial}
\def\d{\mathrm{d}}

\def\jt{\tilde{J}}
\def\jtv{\mathbf{\tilde{J}}}
\def\exp{\mathrm{exp}}
\def\ex{\mathrm{e}}
\def\be{\begin{equation}}
\def\ee{\end{equation}}
\def\bea{\begin{eqnarray}}
\def\eea{\end{eqnarray}}
\def\ie{\textit{i.e.} }

\title{Gauge invariant fluid lagrangian and its application to cosmology}

\author{A. Sulaiman$^{a,b,c}$\footnote{Email : lyman@tisda.org, sulaiman@teori.fisika.lipi.go.id}, \, \, 
T.P. Djun$^a$\footnote{Email : tpdjun@teori.fisika.lipi.go.id} 
\, \, and \, \, 
L.T. Handoko$^{a,d}$\footnote{Email : handoko@teori.fisika.lipi.go.id, laksana.tri.handoko@lipi.go.id}}
\date{}

\begin{document}

\renewcommand{\thesubsection}{\Roman{subsection}}
\maketitle


\thispagestyle{empty}

\begin{center}
\begin{small}
\noindent
$^{a)}$Group for Theoretical and Computational Physics, Research Center for Physics, Indonesian Institute of Sciences\footnote{http://teori.fisika.lipi.go.id}, Kompleks Puspiptek Serpong, Tangerang 15310, Indonesia\\
$^{b)}$P3 TISDA BPPT\footnote{http://tisda.bppt.go.id}, BPPT Bld. II (19$^{\rm th}$ floor), 
Jl. M.H. Thamrin 8, Jakarta 10340, Indonesia\\
$^{c)}$Department of Physics, Bandung Institute of Technology\footnote{http://www.fi.itb.ac.id}, Jl. Ganesha 10, Bandung 40132, Indonesia\\
$^{d)}$Department of Physics, University of Indonesia\footnote{http://www.fisika.ui.ac.id}, Kampus UI Depok, Depok 16424, Indonesia\\
\end{small}
\end{center}

\vspace*{5mm}

\begin{abstract}
A lagrangian for relativistic fluid systems with matters inside is developed using gauge principle. In the model, the gauge boson represents the fluid field in a  form $A_\mu \equiv \epsilon_\mu \, \phi$, where $\epsilon_\mu$ contains the fluid kinematics and $\phi$ is an auxiliary field representing the fluid distribution. This leads to a new relativistic equation of motion for fluid, but which further coincides to the classical Euler equation at non-relativistic limit. The lagrangian is applied to model homogeneous universe as a bulk pure  fluid system. Taking the simplest case of fluid with radial velocity and uniform distribution, the free energy density is calculated and its behaviour around Hubble distance is discussed.
\end{abstract}

\vspace*{5mm}
\noindent
Keyword(s) : astrophysical fluid dynamics, gauge theory, lagrangian

\clearpage

\subsection{Introduction}

Since long time ago, (especially relativistic) fluid dynamics has been applied widely in astrophysics to deal with some fluid like phenomena in our universe \cite{jeans,gibson,gourgoulhon}. This approach is mainly motivated by the lack of interaction-based approach familiar in nuclear and particle physics that leads to in most cases unsolvable many body systems \cite{battaner,clarke}. The relativistic fluid is highly successful to describe the dynamics of many-body relativistic systems. It has been used to model some systems as small as heavy ion collision, and also as large as our universe with intermediate sized objects like neutron stars being considered along the way \cite{andersson}. 

Most of those fluid-inspired models are constructed from the energy momentum tensor $T^{\mu\nu}$ to take into account, not only the bulk motion of fluid, but also for example the random thermal velocity of each fluid particle and various forces between them which should contribute to the total potential energy of system. For instance, in the case of perfect fluid, it takes the well-known form $T^{\mu\nu} = (\rho + P) u^\mu u^\nu - P g^{\mu\nu}$, with density $\rho$, pressure $P$ and $u^\mu \equiv \gamma (1,\v)$ for $\gamma \equiv \left(  1 - \beta^2 \right)^{-1/2}$ ($\beta \equiv {|\v|}/c$). It is then argued that one can derive the continuity equation through the conservation of energy described by $\pd_\mu T^{\mu\nu} = 0$ \cite{landau}. This method is actually analogous to classical fluid dynamics, that is starting from the continuity equation to derive the equation of motion (EOM) or vice versa. However, this kind of approach can not resolve the microscopic dynamics of a macroscopic bulk system modelled as a fluid system. This is crucial in some large scale objects like universe where inhomogeneity might significantly contribute to the whole system.

In contrast with the classical approach, one can in principle construct the quantum fluid dynamics using the principle of quantum mechanics. For example, some works have been done in quantizing fluid either in the Hamiltonian framework using the Clebsch parametrization \cite{ghosh}, or using canonical expressions \cite{batalin,mitra,anishetty}. Although the gauge symmetry has been adopted in these works, the models still require further generalization to extend them to the relativistic ones. There are also some works on gauging the fluid in the classical framework to incorporate electromagnetic interaction in a charged fluid system \cite{bambah}. This kind of magnetohydrodynamics models and its relativistic versions are especially relevant for plasma physics as quark-gluon-plasma (QGP), heavy ion collision \cite{clare} and so on.

Further advancements are adopting the lagrangian method. The main common advantage of this  method is we are able to expand our theory to accomodate more physical interests like vorticity etc in a consistent manner since the initial properties of lagrangian would be conserved along the procedures \cite{asada1,asada2}. Moreover, the gauge extensions along these lines have been done for non-abelian fluid dynamics \cite{jackiw}, and for ideal fluid with global SO(3) symmetry \cite{kambe}. We remark here that all of them have been developed starting from continuity equation. 

On the other hand, inspired by similarities between the Maxwell and fluid theories \cite{marmanis}, the classical fluid equation should be able to be reproduced directly from the electromagnetic lagrangian. However, in this paper we follow rather different approach, that is starting from a predefined gauge invariant relativistic lagrangian, and further deriving the EOM using Euler-Lagrange equation. This would avoid complexities on transforming the equations of non-relativistic to relativistic dynamics, since we start from fully relativistic framework. Moreover, we propose a different form of gauge boson which might also be appropriate for performing numerical calculation using lattice gauge theory later on. 

The paper is organized as follows. First we introduce the model and derive the EOM and continuity equation for relativistic fluid.  Before concluding the paper, we provide a simple example of modelling universe as a pure fluid system in the model, and calculate its free energy density. 

\subsection{The model}

Concerning the basic fluid properties which has no any intrinsic quantum numbers like spin, first let us consider a relativistic lagrangian (density) for bosonic matter,
\be
        \l_\Phi = \left( \pd_\mu \Phi \right)^\ast  \left( \pd^\mu \Phi \right) + V(\Phi) \; , 
        \label{eq:lphi}
\ee
where $V(\Phi)$ is the potential. For example in typical $\Phi^4$ theory, $V(\Phi) = \frac{1}{2} m_\Phi^2 \, \Phi^\ast \Phi + \frac{\lambda}{4} (\Phi^\ast \Phi)^2$. 

Inspired by the quantum electrodynamics (QED) theory, we put the lagrangian to be invariant under local  Abelian gauge transformation \cite{yang,mills}, $U \equiv \exp[-i \theta(x)] \approx 1 - i \theta(x)$ with $\theta \ll 1$. The boson field is then transformed as $\Phi \stackrel{U}{\longrightarrow} \Phi^\prime \equiv \exp[-i \theta(x)] \, \Phi$. It is well-known that the symmetry in Eq. (\ref{eq:lphi}) is revealed by introducing a gauge field $A_\mu$ which is transformed as $A_\mu \stackrel{U}{\longrightarrow} {A_\mu}^\prime \equiv A_\mu - \frac{1}{g}  (\pd_\mu \theta)$, and replacing the derivative with the covariant one, $D_\mu \equiv \pd_\mu - i g \, A_\mu$. Further the gauge invariant kinetic term for gauge boson takes the form of $F_{\mu\nu} {F}^{\mu\nu}$, with strength tensor $F_{\mu\nu} \equiv \pd_\mu A_\nu - \pd_\nu A_\mu$.

Finally the total lagrangian with such gauge symmetry becomes,
\be
	\l = \l_\Phi + \l_A + \l_\mathrm{int} \; ,
	\label{eq:l}
\ee
where,
\bea
        \l_A & = & -\frac{1}{4} F_{\mu\nu} F^{\mu\nu} \; ,
        \label{eq:la} \\
	\l_\mathrm{int} & = & g J_\mu A^\mu + g^2 \left( \Phi^\ast \Phi \right) A_\mu A^\mu \; ,
        \label{eq:li}
\eea
while the 4-vector current is,
\be
        J_\mu = -i \left[ (\pd_\mu \Phi)^\ast \Phi - \Phi^\ast (\pd_\mu \Phi) \right] \; , 
        \label{eq:j}
\ee
which satisfies the current conservation $\pd^\mu J_\mu = 0$ respectively. The coupling constant $g$ represents the interaction strength between gauge field and matter. We note that the second term in Eq. (\ref{eq:li}) gives rise to the mass of fluid field, \ie $m_A^2 \equiv g^2 \left\langle \Phi\right\rangle^2$, 
where  $\left\langle \Phi\right\rangle$ is the vacuum expectation value (VEV) of matter field. 

Rather than following the same argument in \cite{marmanis}, here we construct the fluid lagrangian similar to the QED-like lagrangian as $\l_A$ above, that is considering $A_\mu$ as a ``fluid field'' with velocity $\v$. Further, let us take an ansatz that the gauge boson has the following form,
\be
 	A_\mu = \left( A_0, \a \right) \equiv \epsilon_\mu \, \phi 
	\label{eq:a}
\ee
with, 
\be
        \epsilon_0 = \frac{1}{2} {\gamma}^2 \left| \v \right|^2 
	\; \;  \; \; \mathrm{and} \; \; \; \; 
        \epsilon_i = -\gamma \, v_i \; ,
        \label{eq:ae}
\ee
where $\phi$ is an auxiliary boson field, while again $\gamma \equiv \left( 1 - {\beta}^2 \right)^{-1/2}$ and $\beta \equiv {|\v|}/c$. We should remark that the auxiliary field $\phi$ is introduced in this paper to keep correct dimension. This actually represents the distribution of fluid in the system. Also, this separation is crucial in numerical calculation using lattice gauge theory later on, where we need to split the gauge field and its kinematic parameter to keep the velocity $\v$ appearing throughout the calculation. Because in lattice simulation the gauge boson is normally integrated out completely. However this point is out of coverage of this paper and will be discussed in detail in subsequent paper. 

Using Euler-Lagrange equation in term of $A_\mu$ against Eqs. (\ref{eq:la}) and (\ref{eq:li}), we obtain the following EOM,
\be
   \pd_\mu (\pd^\nu A_\nu) - \pd^2 A_\mu - g J_\mu - g^2 \left( \Phi^\ast \Phi \right)  A_\mu = 0 \; .
   \label{eq:eom}
\ee
However, the last term in Eq. (\ref{eq:eom}) breaks the gauge invariance. This can be resolved by taking the VEV of matter to be $\left\langle \Phi\right\rangle = 0$. So, the EOM becomes $\pd_\mu \, F^{\mu\nu}= g J_\mu$, where non-trivial relations are obtained for $\mu \ne \nu$. We are now ready to derive the EOM relevant for relativistic fluid dynamics. Substituting Eqs. (\ref{eq:a}) and (\ref{eq:ae}) into the EOM we obtain,
\be
	\gamma \, \frac{\pd \v}{\pd t} + \frac{1}{2} {\gamma}^2 \, \left( \grad \left|\v\right|^2 \right) + \v \, \frac{\pd \gamma}{\pd t} + \frac{1}{2} |\v|^2 \left( \grad {\gamma}^2 \right) = -g \jtv \; ,
	\label{eq:ree}
\ee
for uniform $\phi$, \ie $\phi$ : constant. Here, $\jt_i \equiv \int \d x_i J_0 = -\int \d t J_i$. Taking non-relativistic limit, \ie $\beta \rightarrow 0$ ($\gamma \rightarrow 1$), we find, 
\be
  \frac{\pd \v}{\pd t} + (\v \cdot \grad) \v = - (\v \times \om) - g \jtv \; ,
  \label{eq:cee}
\ee
utilizing the vector identity $\frac{1}{2} \grad \left| \v \right|^2 = (\v \cdot \grad) \v + \v \times (\grad \times \v)$, and $\om \equiv \grad \times \v$ is the vorticity. Actually, Eq. (\ref{eq:cee}) reproduces the classical EOM for (turbulent) fluid. Therefore, one can conclude that Eq. (\ref{eq:ree}) should be its relativistic version. Moreover, the lagrangian in Eq. (\ref{eq:l}) with $A_\mu = \epsilon_\mu \, \phi$ written in Eq. (\ref{eq:a}) should describe a general relativistic fluid system interacting with the matters inside.

We should remark few points here. First, the additional ``current force'' in Eqs. (\ref{eq:ree}) and (\ref{eq:cee}) is induced by the external source as manifestation of interacting fluid with the matters inside. Secondly, in the classical fluid dynamics the 4-vector current $J_\mu = (\rho, \rho \v)$ represents the macroscopic distributions of fluid density (charge) and current. On the other hand, in our approach $J_\mu$ reveals the dynamics of the distribution function of matters inside the fluid as expressed in Eq. (\ref{eq:j}). This is analogous to the current in QED which is the result of fermion-pair interaction by emitting photon, and not the current of electromagnetic field itself as in the Maxwell theory. 

Now we are going to apply the lagrangian to model the universe. We should emphasize that using the lagrangian would benefit us from calculating relevant physical observables without dealing with unsolvable nonlinear EOM of fluid as shown in the subsequent section. Especially for gauge invariant lagrangian, we can further make use of known techniques in lattice gauge theory to perform any relevant calculations non-perturbatively. This would avoid unnecessary assumptions to enable perturbative calculations, since we should not in general assuming that $g$ is small enough in the present model. 

\subsection{Example : pure and homogeneous fluid universe}

Let us consider about modelling a homogeneous universe without any matter inside as a relatisitic fluid system. This means the matter and interaction terms in the lagrangian can be omitted, and we take into account only the pure gauge term, that is $\l = \l_A$. 

In order to calculate, from the lagrangian, relevant physical observables in a system with finite temperature $T$ in a unit volume, we bring the partition function density, 
\be
	\z = \ex^{-\frac{1}{T} \h} \; ,
	\label{eq:z}
\ee
where $\h$ is the hamiltonian density. Using $T_{\mu\nu} = \left[ {\pd \l}/{(\pd(\pd^\mu A_\rho))}\right] \pd_\nu A^\rho - g_{\mu\nu} \l$, we have the energy-momentum tensor being $T_{\mu\nu} = -(\pd_\mu  A^\rho) \, F_{\rho\nu} \,  + \frac{1}{4} g_{\mu\nu} \, F_{\rho\sigma} F^{\rho\sigma}$. However, this is not symmetric and gauge invariant. The conserved, symmetric and gauge invariant $T_{\mu\nu}$ should be $T_{\mu\nu} = -F_{\mu\rho} {F^\rho}_\nu + \frac{1}{4} g_{\mu\nu} \, F_{\rho\sigma} F^{\rho\sigma}$ by adding a term $(\pd^\rho A_\mu) F_{\rho\nu}$. This additional term is admissible since in the case under consideration, no matter inside the fluid such that the current in Eq. (\ref{eq:j}) is absence. This leads to $\pd^\rho F_{\rho\nu} = 0$ and then $\pd^\rho (A_\mu) F_{\rho\nu} = \pd^\rho (A_\mu F_{\rho\nu})$. Because the hamiltonian density is nothing else than the $00-$element of $T_{\mu\nu}$, we obtain, 
\be
	\h = \frac{1}{2} \left[ \left| \frac{\pd \a}{\pd t} + \grad A_0 \right|^2 + \left| \grad \times \a \right|^2 \right]  \; ,
	\label{eq:h}
\ee
which is also already known in the electromagnetic theory as the electric and magnetic terms. Assuming that the fluid has radial velocity, \ie $\v = v(t,r)$ in polar coordinate, we calculate for each term,
\bea
	\left. \frac{\pd \a}{\pd t} + \grad A_0 \right|_r & = & -\gamma \left( \frac{\pd v}{\pd t} - v \gamma \frac{\pd v}{\pd r} \right) \, \phi \; ,
	\label{eq:e}\\
	\left. \grad \times \a \right|_r & = & 0 \; ,
	\label{eq:b}
\eea
by substituting $A_\mu$ in Eqs. (\ref{eq:a}) and (\ref{eq:ae}), and putting uniform fluid distribution ($\phi$ : constant) in consequence of homogeneous fluid. 

The free energy density is given by $\f = -T \, \ln \z$. Therefore in the present case it yields,
\be
	\f = \frac{1}{2} v^2 \gamma^4 \left( \frac{\pd v}{\pd r} \right)^2 \, | \phi |^2  \; , 
	\label{eq:f}
\ee
for steady velocity $v(t,r) = v(r)$ using Eqs. (\ref{eq:z}), (\ref{eq:h}), (\ref{eq:e}) and (\ref{eq:b}). It is straightforward to see that the free energy density is being infinite as $\beta \rightarrow 1$. In other words, borrowing the Hubble law, $v = r H$ \cite{hubble}, we can deduce from Eq. (\ref{eq:f}) that,
\be
	\f \rightarrow \infty \; \; \; \; \mathrm{at} \; \; \; \; r \rightarrow r_\mathrm{th} = c \, H^{-1} \; .
	\label{eq:resultf}
\ee
This threshold distance $r_\mathrm{th}$ is nothing else than the Hubble distance. 

\subsection{Summary and discussion}

We have introduced a model for fluid universe with interacting matters inside.  The lagrangian has been constructed using gauge principle and taking particular form of gauge boson as Eq. (\ref{eq:a}). The lagrangian should provide general description of relativistic fluid dynamics. This has been proven in the  case of Abelian lagrangian with  uniform field distribution which reproduces the classical Euler equation for turbulent fluid at non-relativistic limit. Using the lagrangian, we have calculated the free energy density for non-interacting homogeneous universe. In the present model the free energy would be getting infinite as approaching the Hubble distance, \ie the universe horizon, where the system is completely relativistic. Beyond the horizon, the entropy is revived to be finite. 

According to this result, one might speculate that the horizon is like a mirror of twin universe if the fluid distributions in both sides are completely same. However one might take another  functions for $\phi$ with some reasons. For instance, if $\phi$ follows a step function,
\be
	\phi = \phi(r) = \left\lbrace \begin{array}[3]{lcl}
		1 	& , & r < r_\mathrm{th} \\
		\ex^{\alpha r}		& , & r \geq r_\mathrm{th} 
	\end{array}
	\right.  \; ,
\ee
with a constant $\alpha$, the free energy beyond the horizon would be suppressed close to null for a quite large $\alpha$. 

In contrary, at non-relativistic limit ($\beta \rightarrow 0$) the free energy should be proportional to the square of velocity. This result coincides with the classical kinetic theory of gas. 

\vspace*{5mm}
\noindent
\textbf{Acknowledgement}

We greatly appreciate fruitful discussion with N. Riveli, H.B. Hartanto and A. Oxalion throughout the work. AS thanks the Group for Theoretical and Computational Physics LIPI and the Bandung Institute of Technology, while TPD thanks the Group for Theoretical and Computational Physics LIPI for warm hospitality during the work. This work is partially funded by the Indonesia Ministry of Research and Technology and the Riset Kompetitif LIPI in fiscal year 2007 under Contract no.  11.04/SK/KPPI/II/2007.

\end{document}